# Growth-related formation mechanism of I₃-type basal stacking fault in epitaxially grown hexagonal Ge-2H

L.Vincent[a]*, E.M.T. Fadaly[b], C. Renard[a], W.H.J. Peeters[b], M. Vettori[b,] F. Panciera[a], D. Bouchier[a], E.PA.M Bakkers[b], M.A. Verheijen[b]

a Université Paris-Saclay, CNRS, Centre de Nanosciences et de Nanotechnologies, 91120, Palaiseau, France

b Department of Applied Physics, Eindhoven University of Technology, Groene Loper 19, 5612AP Eindhoven, The Netherlands

*Corresponding author: laetitia.vincent@c2n.u-psaclay.fr

**Abstract**

The hexagonal-2H crystal phase of Ge recently emerged as a promising direct bandgap semiconductor in the mid-infrared range providing new prospects of additional opto-electronic functionalities of group-IV semiconductors (Ge and SiGe). The controlled synthesis of such hexagonal (2H) Ge phase is a challenge that can be overcome by using wurtzite GaAs nanowires as a template. However, depending on growth conditions, unusual basal stacking faults (BSFs) of I₃-type are formed in the metastable 2H structure. The growth of such core/shell heterostructures is observed *in situ* and in real-time by means of environmental transmission electron microscopy using chemical vapour deposition. The observations provide direct evidence of a step-flow growth of Ge-2H epilayers and reveal the growth-related formation of I₃-BSF during unstable growth. Their formation conditions are dynamically investigated. Through these *in situ* observations, we can propose a scenario for the nucleation of I₃-type BSFs that is likely valid for any metastable hexagonal 2H or wurtzite structures grown on m-plane substrates. Conditions are identified to avoid their formation for perfect crystalline synthesis of SiGe-2H.

**Graphical abstract**

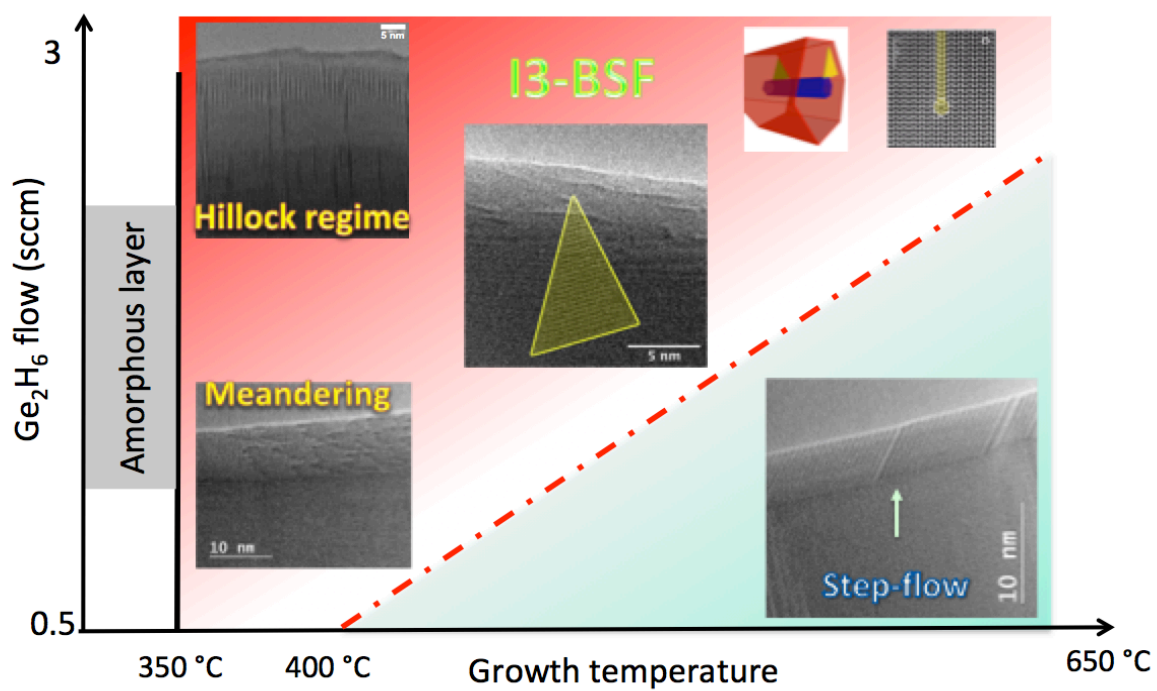

**Keywords :**

Kinetics growth

Polytypism

Epitaxy

Stacking faults

Step-flow

# 1. Introduction

Polytypism of semiconductors is explored to achieve new heterostructure configurations and modify electronic and photonic properties for additional functionalities, such as crystal phase quantum dots,[1-2] and tuning the magnitude and the nature of the semiconductor band gap.[3] The stable crystal phase of Si and Ge is the well-known cubic diamond structure (herein noted 3C) corresponding to the lowest total energy.[4, 5] The diamond hexagonal structure 2H has the second lowest energy;[6] the difference in total energy is around 0.015 eV/atom for both Si and Ge. [4] By epitaxy on a hexagonal structure with a suitable lattice parameter, one can overrule this energy configuration in defiance of the 3C bulk stability to obtain the 2H phase. This was first demonstrated with Si/GaP core/shell heterostructures grown by Metal Organic Chemical Vapor Deposition (MOCVD) as a Si-2H structure was epitaxially grown on the sidewalls of a <0001>-oriented wurtzite (WZ) GaP nanowire.[7, 8]

For Ge-2H, the ideal template is wurtzite GaAs (GaAs-WZ) because their lattice constants are almost identical ($\Delta a = 0.03\%$ and $\Delta c = 0.11$ %).[9] While GaAs bulk has a zinc-blende (ZB) structure, nanowires grown by the vapor-liquid-solid process may exhibit a WZ structure depending on growth parameters.[10-12] Thus, control over the GaAs-WZ phase in nanowires gave a decisive opportunity to transfer the structure to a Ge shell.[9] This crystal structure transfer method enabled the synthesis of $Si_{1-x}Ge_x$-2H shells from $x = 1$ down to $x = 0.6$, which exhibit direct band gap emission.[9]

For both core/shell systems, Si-2H/GaP-WZ and Ge-2H/GaAs-WZ, unusual $I_3$-type basal stacking faults ($I_3$-BSF) are frequently observed, [13, 14]. Although $I_3$-BSF in $Si_{1-x}Ge_x$-2H may have no deleterious influence on optical properties as reported by Fadaly *et al*. their presence may affect the electrical and mechanical behaviour of the material.[14] Such defects are intrinsic faults with a characteristic sequence ABA**BACAB**ABA where only one C-plane replaces a B-plane (or an A-plane in the sequence BABABCBABA). On both sides of the C-plane the crystal remains perfect *i.e.* unshifted (no associated Burgers vector). The configuration of the $I_3$-BSF is almost never found in the hexagonal system, although it was theoretically predicted in nitrides as a growth-related defect with the second-lowest formation energy just above that of $I_1$-BSF.[15] We think that similar defects are observed in InAs/GaAs core/shell structures but their nature needs to be confirmed.[16] Indications of this unusual defect have been found in GaN grown on (11-20) 4H-SiC and in the defected tip of GaAsP nanowires.[17-19] High resolution STEM imaging showed a change in the orientation of the dumbbells in the (0001) planes with a null total Burgers vector.[18] Thus far, an explanation for the formation mechanism of this defect has not been proposed.

Here, we study the formation mechanism in detail, using *in situ* and *ex situ* experiments, in order to optimize the synthesis of high quality 2H polytype in SiGe compounds. Firstly, the *ex situ* growth experiments presented in ref [14] are extended here by further exploring the defect shape and the effects of both temperature and shell volume on the defect density. We also assess the influence of the chemical surface treatment for removing gold from the GaAs NW prior to the Ge shell growth. In these experiments, structural characterisation is performed by means of post-growth transmission electron microscopy (TEM) studies. Secondly, with *in situ* environmental transmission electron microscopy, we monitor the growth of Ge-2H and we examine the parameters influencing the defect nucleation. This *in situ* TEM study is carried out using the NANOMAX microscope (http://www.tempos.fr/?q=node/3) implemented with gas precursors for chemical vapor deposition growth. This microscope provides real-time and atomic-level information about the growth dynamics, useful to understand involved mechanisms in growth and defect formation. Real-time observations give direct insight into the different growth modes depending on temperature and precursor flows. These observations reveal a correlation in the growth conditions under which the destabilisation of the step-flow growth and the nucleation of a $I_3$-BSF occur. The results provide us with convincing arguments to propose scenarios for the formation of $I_3$-BSF during the growth of metastable hexagonal structures.

## 2. Theory

Before we start discussing the experimental results, we give the structural characteristics of the $I_3$-BSF in Ge-2H. Figure 1a shows a schematic view of a representative core/shell GaAs-WZ/Ge-2H structure featuring the crystallographic directions. The GaAs-WZ nanowire is grown along the <0001> axial direction and displays prismatic {1-100} sidewalls. High resolution STEM micrographs acquired along the <11-20> zone axis enable to identify the nature of the defects present in these wires. Lamellar twins with an $I_1$ or $I_2$ stacking fault configuration may be formed during the axial growth of the GaAs nanowires corresponding to a local ZB structure. Those defects are consequently replicated by epitaxy in the Ge shell. Their density can be minimized by tuning the GaAs growth parameters to obtain long GaAs-WZ cores. Here, for ex situ analyses, we use GaAs-WZ core nanowires with around 5 SFs/ $\mu$m. Besides these $I_1$ or $I_2$ stacking faults, unexpected intrinsic $I_3$-BSFs are observed as dark lines starting and extending across the Ge-2H shell on the basal (0001) planes (Fig. 1b). These I$_3$ BSFs do not appear in the GaAs cores as confirmed in the supporting information in ref. [14].

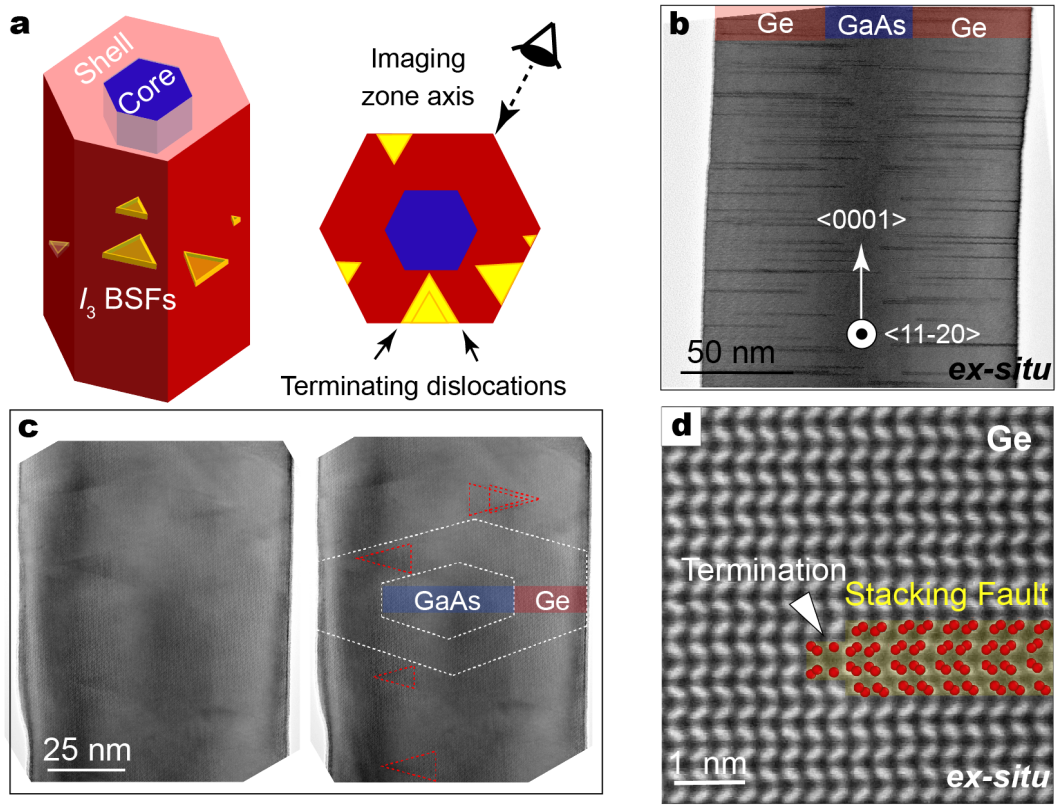

**Figure 1. a)** Schematic illustration of I$_3$ defected GaAs-w/Ge-2H core/shell structure grown in <0001> direction. **b)** BF-TEM image, imaged along the <11-20> zone axis, of an ensemble of I$_3$-BSFs in the Ge-2*H* indicated by partial dark lines running across the Ge shell. The relevant crystallographic directions are indicated on the image. **c)** The triangular shape of the I3 BSF is resolved under a tilted view of ~32° away from the <11-20> zone axis, i.e. along the <-2112> direction yielding a triangular area with moiré fringes (zoom in panel). The red lines highlight the boundaries of the I3 BSF (triangles, of the hexagonal core and of the planar stacking fault over the full diameter of the wire. d) HAADF-STEM image showing the ABAB stacking of the 2H Ge shell and the ABCBA stacking of the I3 BSF stacking, highlighted in red.

A detailed description of the atomic stacking within these particular defects is also given in ref. [14]. Here, we summarize the foremost structural characteristics of the I$_3$-BSFs in the Ge-2H shell based on TEM analyses (Fig. 1), providing insights in the stacking structure, the boundary of the faulted layer and the shape of the I$_3$-BSF:

i) Along the [0001] direction, the defective sequence ABA**C**ABA is equivalent to a local twinned cubic structure with only one faulted bi-atomic layer. There is no glide of the upper and lower part of the 2H crystal (Fig. 1d). The defects are mostly constituted of one faulted plane, we rarely observe two or more faulted planes

ii) Viewing along a <11-20> direction, one can define a topological defect with a translation of two successive monoatomic layers (forming two partial 30° dislocations on the glide plane) resulting in a change in the orientation of the dumbbells on either side of the faulted (0001) plane with respect to the dumbbells on the perfect hexagonal stacking. The core of the defective boundary shows two vacancy rows bounding the two translated monoatomic layers (see Fig. 1c).

iii) With a tilted view away from the <11-20> viewing direction, the defects are visible with moiré fringes due to the overlay of one faulted plane with the undefected lattice (Fig.1c). The moiré pattern is localised in a restricted area. The majority of defects exhibits a triangular shape with one side lying on the growth surface and the opposite apex corresponding to the nucleation point of the defect inside the bulk of the Ge-shell. In the basal plane, the resulting equilateral triangle is bounded by two defective lines along <11-20> directions, indicating a point source of the defect nucleation at the triangle apex. This equilateral triangular shape implies that the maximum projected length of the defect (*i.e.* the triangle height) observed along a <11-20> direction is equal to the shell thickness and can thus be identified using a <11-20> zone axis view. It is important to note that these triangular defects are not necessarily formed at the GaAs/Ge interface. They may start at random positions in the shell and most of them have thus length smaller than the shell thickness.

i) Additionally, we have occasionally observed few elongated defects with a projected length larger than the shell thickness and a shape different from a triangle (results of elongated defects are not shown here). They are still bounded by <11-20> defected lines, but they do not have a point origin. They are always anchored on the GaAs interface along a <11-20> row and they are never found to be created within the shell, conversely to triangles.

It appears that both defects, *i.e.* with triangular and elongated shapes, have a different starting configuration: at a point and a line, respectively. After nucleation both defects grow bounded by <11-20> dislocation lines. Elongated defects are always formed at the interface and thus their density does not evolve with the growth while triangle defects are randomly formed in the Ge shell during growth. The study in this manuscript focuses on the triangular defect formation. A follow-up paper will discuss separately in detail the elongated defects which may have a different origin.

## 3. Results

We start by studying *ex situ* the kinetics of $I_3$-BSF nucleation by quantitative consideration of their density evolution with the growth time *i.e.* with the shell volume and the growth temperature (information for the calculation of their density was provided in supporting information of ref [14]). As shown in Fig. 2a, at all growth temperatures, the linear dependence of density with volume indicates that the nucleation probability scales with the expanded growing surface. Together with the random position of the nucleation points, at least for the concerned triangular defects, this linear density is consistent with a growth-related defect formation dismissing the hypothesis of a strain relief due to thermal expansion. The plot of Fig.2a also evidences a higher nucleation rate at lower temperatures. The intersection of each linear curve with the y-axis gives the initial $I_3$ density at the start of the growth, *i.e.* at the GaAs/Ge interface. At low growth temperatures, a finite $I_3$ density is present from the beginning of the shell growth while at higher temperatures the curve may intersect the x-axis implying a defect free start of the shell growth. The use of a high growth temperature seems to hinder or delay the nucleation of $I_3$-BSFs.

In Fig. 2b, the effect of temperature on the defect density is plotted in the range 405 (low temperature)-565°C (high temperature) for two different GaAs core surface conditions; one is an as-grown nanowire with the presence of the Au catalyst particle on top and the second is an analogous nanowire treated to etch away the Au catalyst particle. For both, the $I_3$-defect linear density is inversely proportional to the temperature. The shells grown on chemically treated nanowires are more defective than those grown on the untreated, as-grown GaAs wires yet showing no Au contamination. Remarkably, the TEM micrographs (Fig. 2c) show that in the chemically treated NW the defects are mostly anchored at the core/shell interface. Thus, a chemical etching of the GaAs template may induce deleterious surface effects such as roughness or contamination that enhances the nucleation of $I_3$ defects at the interface. However, the convergence of the density curves at high growth temperature suggests an improvement of the GaAs surface quality, either by desorption of chemical contaminants or by evaporation of GaAs, thereby flattening the sidewall surfaces. Actually, we find that the deleterious effect of the chemical etching of the GaAs nanowire can be restored by thermal annealing under $AsH_3$ pressure before Ge shell growth; the Ge growth is carried out using a dedicated Ge reactor lining, such that the As background pressure is low during Ge growth.

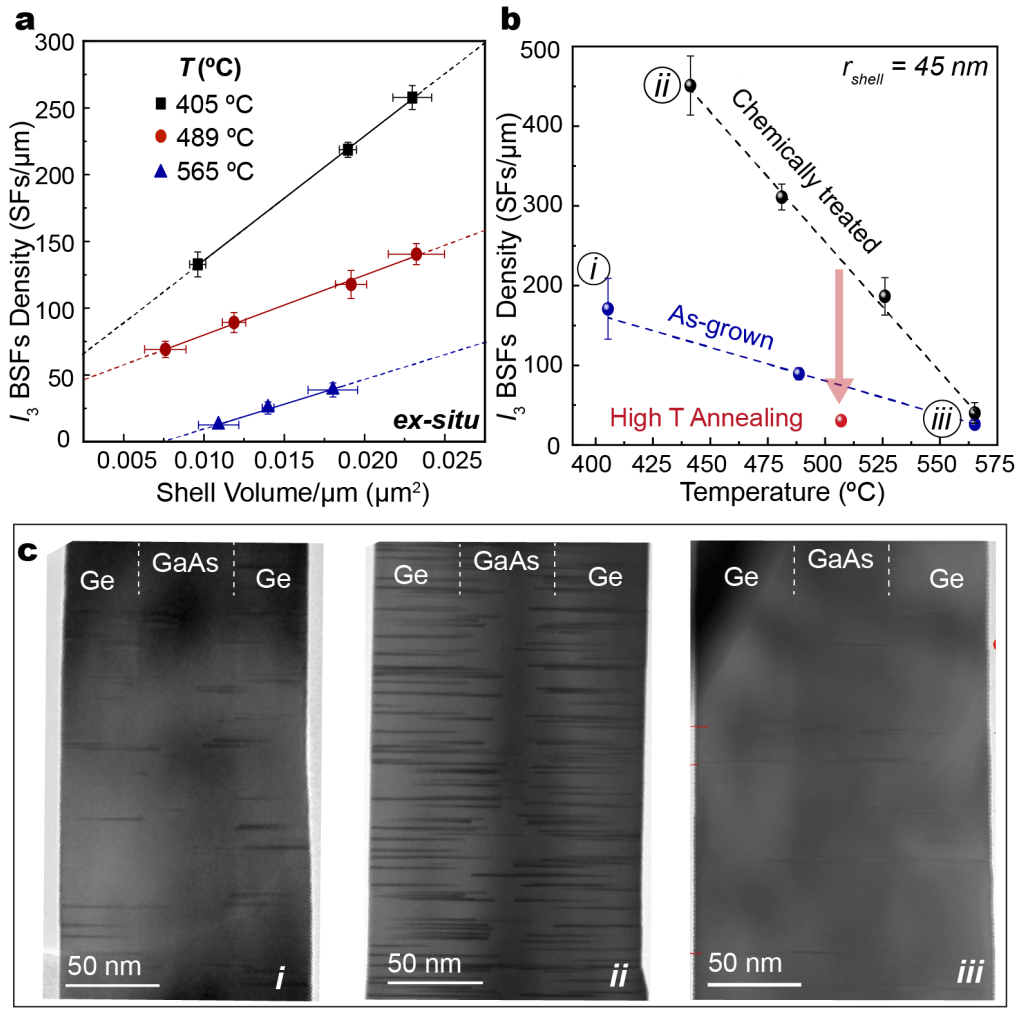

**Figure 2. a)** $I_3$-BSFs density evolution as a function of the substrate surface temperature for various conditions of pre-shell growth surface treatments **b)** $I_3$ BSFs density evolution as a function of the shell volume for different growth temperatures. **c)** BF-TEM images of three different samples : sample(i) grown at relatively low substrate temperature with as-grown GaAs-*w* cores without any pre-growth surface treatments, sample (ii) grown at relatively low temperature with the GaAs-*w* cores chemically treated prior to the Ge-2*H* shell growth, and sample (iii) grown at high temperature, as indicated in panel (a).

As a first statement, roughness and/or impurities may have an influence on the defect nucleation at the interface and growth conditions such as temperature appear to influence the possibility of formation of defects. To minimize the formation of defects in the Ge-2H shell, the highest possible growth temperature should be used. In practice, this implies a temperature around 650°C, as at higher temperatures As evaporation under vacuum conditions affects the GaAs core quality and morphology.[20]

### 3.1 In situ growth-related I$_3$-BSF

To understand the process of the I$_3$-BSF formation, we need to study the conditions under which they are formed during growth and gain insights into the parameters affecting their density. On top of this, we aim to identify the origin of their nucleation. For this goal, *in situ* electron microscopy provides unique opportunities. Starting on the WZ GaAs NW template, after an incubation time that likely depends on temperature, Ge grows on {1-100} sidewalls by perfectly replicating the hexagonal structure of the GaAs-WZ core (Fig. 3a and video SV2), which is thermodynamically most favourable. However, depending on growth conditions, especially when the Ge flux is increased, I$_3$-BSFs appear with increasing number. This may lead to a very defective structure as shown in video SV3 and figure 3b. This real-time observation confirms that such defects are growth-related and are definitely not due to a thermal expansion coefficient mismatch.

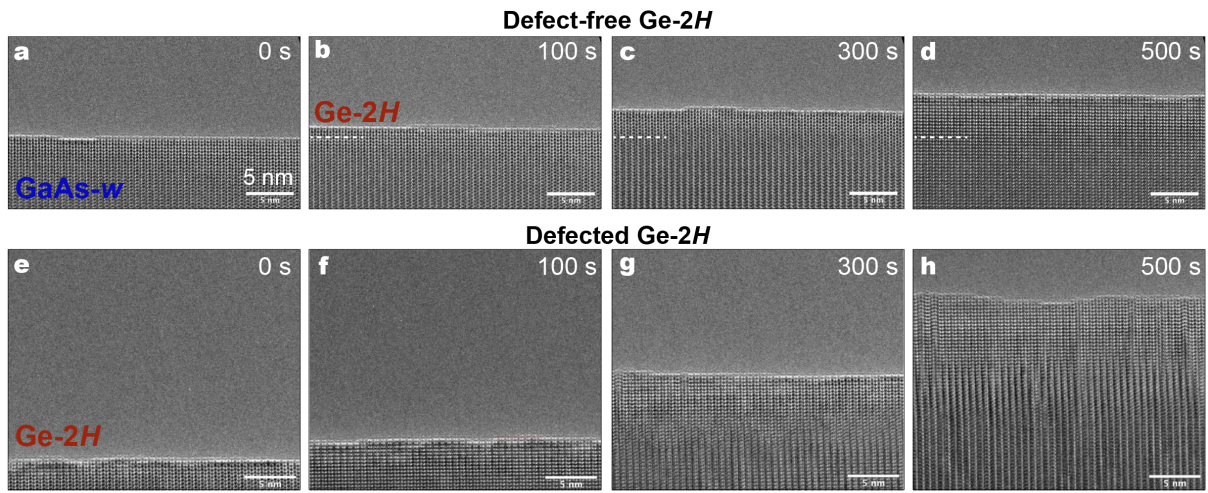

**Figure 3.** TEM micrographs recorded along the <11-20> zone axis during *in situ* growth. **a)** starting from a GaAs-*w* core nanowire template, defect free Ge-2H is grown on m-plane sidewalls at a temperature around 450°C with 1 sccm of Ge$_2$H$_6$/H$_2$ (1/10) and a pressure of around 2x10$^{-3}$ mbar. The growth rate is about 1.5 Å/min. b) Defected Ge-2H after growth at 450°C with 2.5 sccm diluted Ge$_2$H$_6$ (P = 5.3x10$^{-3}$ mbar). I3-BSFs are formed during growth. The growth is observed in real-time as shown in supporting videos SV2 and SV3. The density of *I$_3$*-BSFs increases during the growth (as the shell volume increases). Scale bar is 5 nm

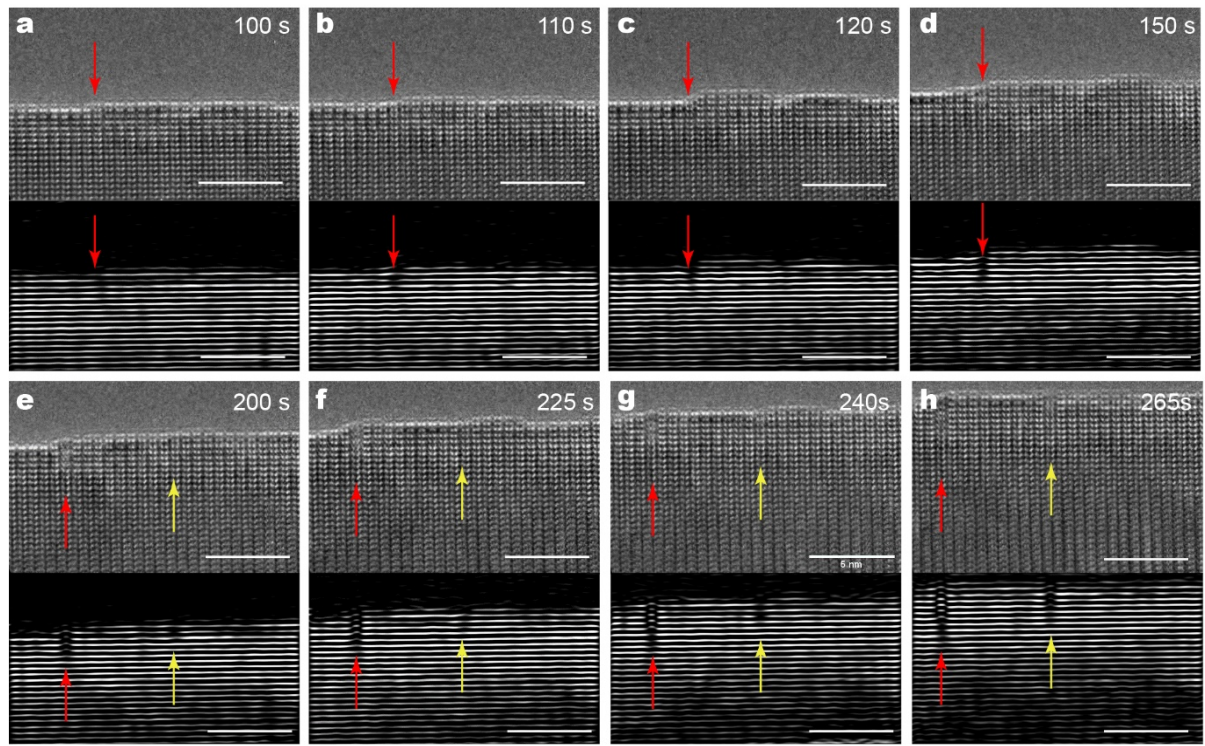

**Figure 4.** HR-TEM images real-time recorded at different times during the growth of Ge-2*H*, as shown in supporting video SV3 and their corresponding filtered IFFT constructed using the (1-100) spatial frequencies in the FFT. IFFT images show the evolution of the $I_3$-BSFs attested by the fringe shifting, as indicated by the red and yellow arrows. Every defect is pointed at with the same color. The scale bar is 5 nm.

With the sequential micrographs in Fig. 4 we can follow the formation and the evolution of an isolated $I_3$-BSF, although it does not allow us to see the exact origin of the defect due to the projection view. However, this particular defect exhibits a C basal plane in place of a A basal plane in the sequence ABABA**C**ABAB along <0001> resulting in a local distortion of the position of the {1-100} planes. This distortion can be visualized by constructing the inverse filtered Fourier transforms (IFFT) of the atomic resolution TEM images. In figure 4, the IFFT corresponding to the HRTEM micrograph acquired at 100 s displays the appearance a dark contrast indicated by a red arrow. This dark area becomes more and more clear with time and with additionally grown Ge layers indicating a small perturbation in the column of atoms. At 120 s the (1-100) fringes in the IFFT are locally cut implying there is a discontinuity of the (1-100) planes. After 150 s, a large bending of the fringes is clearly discernable with a further increased contrast signifying the presence of a $I_3$-BSF. No strain field is observed in the surrounding planes before and after the appearance of the $I_3$-BSF and there is no dissociated dislocation. At 225 s, a second $I_3$-BSF is observed indicated with a yellow arrow. The additionally grown Ge (1-100) planes will continuously display this bending with respect to the other perfectly aligned fringes. Thus, we can see a slight distortion of the (1-100) fringes in the IFFT before it is possible to see the faulted stacking sequence in the image. These successive snapshots reveal that a local perturbation in the (1-100) planes may provoke the formation of the $I_3$-BSF at random positions during growth. As soon as the $I_3$-BSF nucleates, the distortion of the (1-100) planes is replicated on the next planes.

These defects appear randomly with increasing number and size as the growth proceeds. During real-time experiments, we could observe $I_3$-BSF starting at the very beginning of shell growth, especially in the case of low temperature and high $Ge_2H_6$ flow (see SI4). Their formation can be delayed to a later stage in thick shells if we grow the Ge shell at low flow, and I3 appear after an increase of $Ge_2H_6$ flow. This is consistent with the plot in Fig. 2a, showing an increased defect density at the interface for lower growth temperature. This means that $I_3$-BSFs are not related to strain accumulation and their appearance is not forced by a thickness threshold of the Ge shell. This is expected given the very small lattice mismatch between GaAs-WZ and Ge-2H.

During all the real time experiments, we have clearly observed that the density of $I_3$-BSFs increases with decreasing temperature and/or increasing precursor flow. Conditions of low temperature and high flow lead to very defective Ge-2H as illustrated in Fig. 3b. The density of $I_3$-BSF can be reduced using a low precursor flow. At high temperature, their formation is prevented if the flow is not too high and in that case the growth rate does not have any influence (see SI5). Thus, these observations are in agreement with the *ex situ* quantitative analyses of the growth temperature influence on the $I_3$ density.

In addition, the effect of certain chemical impurities on the defect formation has been studied *in situ* by adding precursors such as TMGa or TBAs or $SiH_4$ during growth. Throughout our observations, none of those added precursors have shown any effect on the formation of $I_3$-BSFs in the Ge-shell. Moreover, by comparing qualitatively the shell growth on Au catalysed and self-catalysed NWs, SFs appear to be equally formed on both type of NWs. Thus, there is no evidence for Au, As, Ga, and Si impurity atoms to have any effect on the growth surface that could directly provoke the nucleation of an $I_3$ defect. This point will be discussed further below.

Comparing figures 3a and 3b, in addition to the presence of $I_3$-BSFs we observe a striking difference in the roughness and growth modes. This observation suggests it is essential to control the surface roughness on the nano-scale as a prerequisite to gain understanding in the crystal growth kinetics and the $I_3$-BSF defect formation mechanism.

### 3.2 Growth modes of Ge-2H

The step-flow regime for the growth of planar films has been studied for several decades and has been directly observed by various means such as optical microscopy, reflection electron microscopy, scanning probe microscopy and also TEM, for different film/substrate systems.[21-26] However, our system Ge-2H/GaAs-WZ is singular in the sense that we grow a metastable crystal phase, which opens up new questions about phase stability and possible defect formation during step growth. Here, we describe the first qualitative observations to draw a global picture of the growth process of Ge-2H on m-plane surfaces.

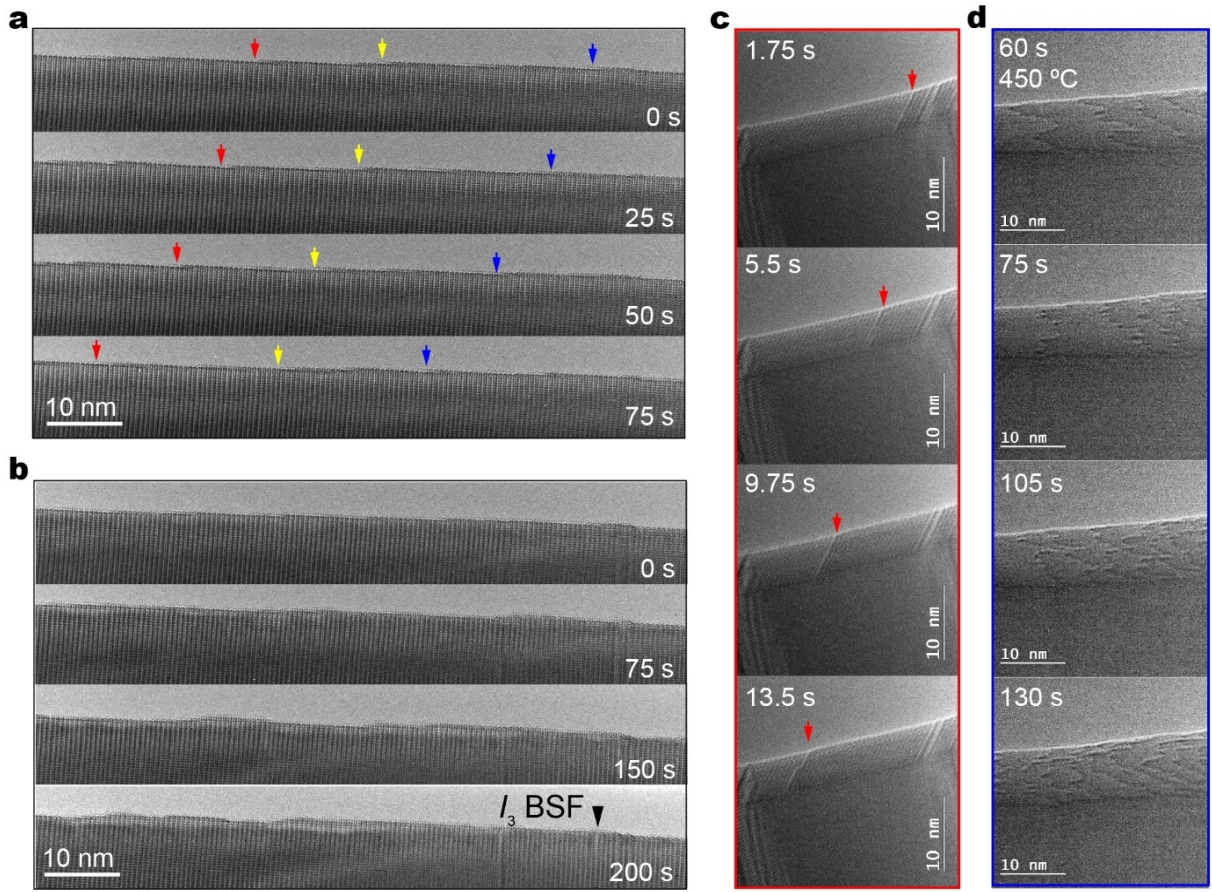

**Figure 5. a)** Snapshots over the growth time of Ge-2*H* from video SV4 at a temperature around 500°C, flow of 1sccm of diluted GeH$_6$ and pressure of 2.5 x 10$^{-3}$ mbar. The step flow growth mode is followed on the (1-100) sidewall viewed along the <11-20> zone axis. Arrows point out the propagation of 3 monolayer steps. The step velocity is of 3 A/s. **b)** Snapshots of Ge-2H growth from video SV8. Evolution of the growth surface after an increase at t$_0$ = 0 s of the deposition flow from 1 to 2 sccm and P = 4.3 x 10$^{-3}$ mbar at T = 500°C. The step flow on the (1-100) surface turns to an irregular and rough surface with the formation of hillocks visible with bright and dark contrast as well as mound on the edge. **c)** The red panel shows snapshots extracted from supporting video SV5 (T= 550°C, 1.2 sccm of diluted $Ge_2H_6$ and P = 1.5 x$10^{-3}$). The step-flow on the (1-100) plane is visible with a regular moving contrast using a tilted view. The monotomic straight step indicated by the red arrow progresses in the <0001> direction. **d)** The blue panel exhibits the evolution of the step meandering due to a decrease of temperature from 450°C to 420°C at $T_0$ using 1sccm of diluted $Ge_2H_6$, P = 1x$10^{-3}$ mbar. The panel is extracted from supporting video SV6.The time is indicated in seconds for each sequential image.

Videos SV4 - SV5 and extracted snapshots in Fig. 5a and 5b reveal a step-flow growth mode favourable for the epitaxy of Ge-2H on the GaAs-WZ with standard conditions of high temperature (500-550°C) and and low partial pressure in the range 1-3 x$10^{-3}$ mbar with a flow around 1 sccm. In Fig. 5a, the overview of the growth surface, observed along the <11-20> zone axis, *i.e.* when the electron beam is parallel to one of the {1-100} growth surfaces, shows several successive steps indicated by arrows with a nearly regular advancement. The bi-atomic layer step (consisting of a dumbbell of Ge atoms) advancement can be even more easily followed with a tilted view on the {1-100} m-plane (snapshots of SV5 in figure 5c), revealing additional information on the advancement of the step. When the growth proceeds, straight steps move one after another over the NW sidewalls. In the <0001> direction, the step front progresses "row-by-row". Adatoms are incorporated at the kink of the row to finish a step before starting a new one, the step proceeding from one edge to the other on the {1-100} sidewall. Thus, the hexagonal m-plane surface is replicated layer-by-layer, step by step and we can say row after row.

Cubic segments initially existing in the GaAs core are also reproduced in the otherwise perfect Ge-2H shell. The nucleation of the step seems to be preferentially initiated at the surface position of the local cubic structure. The enhanced probability of nucleation at these positions might be either due to a lower nucleation barrier for the cubic phase, or to the altered atomic arrangement at the hexagonal-to-cubic stacking sequence. There is no preferential direction of the flow along a NW *i.e.* the step-flow can move upward or downward the NW and is not related to a miscut of the surface. Remarkably, in case of initial sidewall roughness of the GaAs NW (see SI6), this step-flow growth mode favours a surface flattening that may be explained by surface diffusion of adatoms and higher sticking probability at the step edges of lower layers (*i.e.* chemisorbed Ge atoms complete the lower layers).

The growth rate can be tuned by the precursor flow and in a lesser extent by the temperature which may also modify the precursor cracking. Remarkably, decreasing the growth temperature and/or increasing precursor flow induces instabilities of the step-flow of Ge-2H such as step meandering (Fig. 5d and video SV6). Starting from a slightly destabilized step-flow (with 1sccm of $Ge_2H_6$ at 450°C) the snapshots in Fig. 5d show how the growth evolves towards more pronounced step meandering upon lowering the growth temperature to 420°C. Consequently, the steps become highly irregular in shape as well as in advancement speed. A large destabilisation leads to a change in growth mode as illustrated in Fig.6. Keeping precursor flow and pressure constant and decreasing the growth temperature by more than 100°C, the growth mode evolves promptly from step-flow to the formation of hillocks and 2D-nucleation along with the nucleation of $I_3$-BSFs.

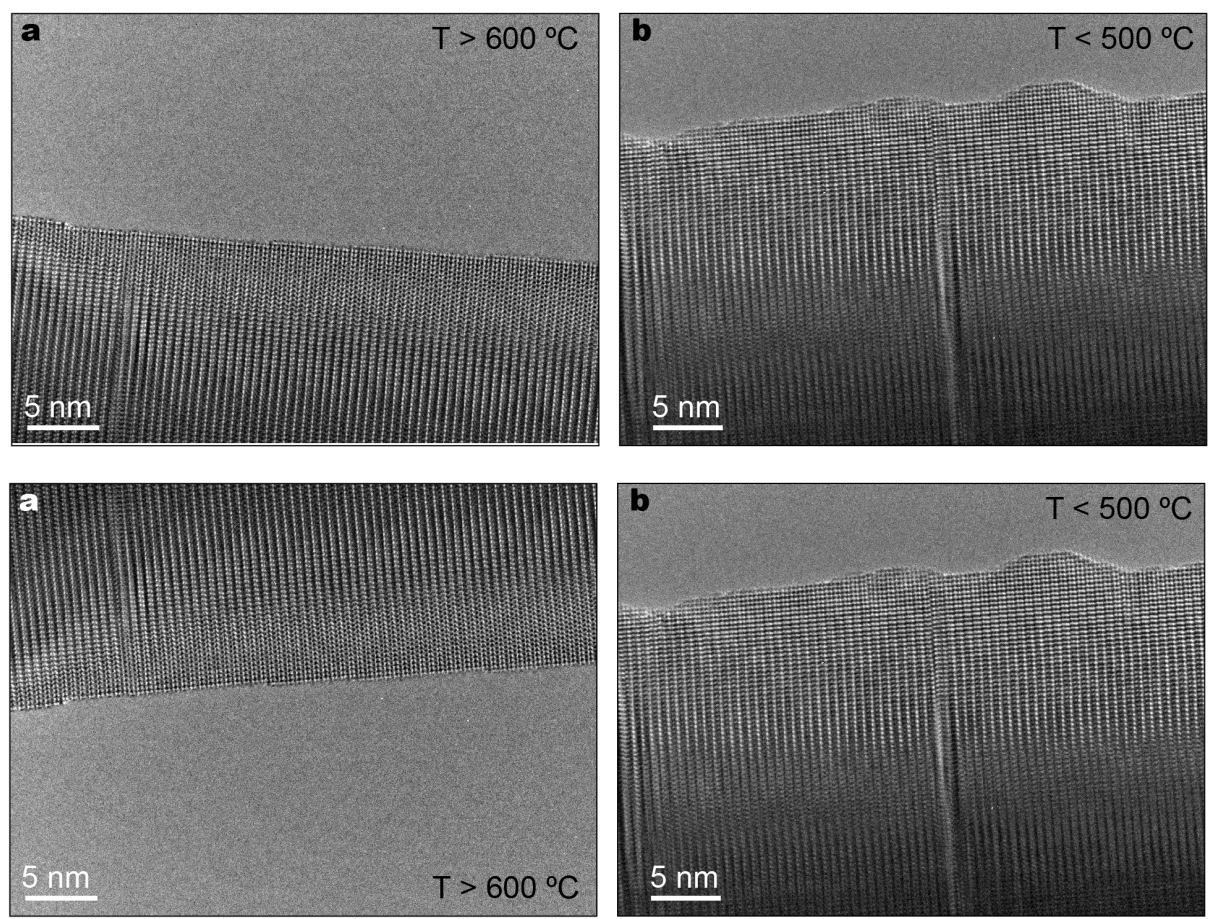

**Figure 6 .** BF-TEM images viewed along the <11-20> zone axis showing the growth of Ge-2H grown at different temperatures : **a)** T > 600 °C, and **b)** T < 500 °C. Growth is performed using 1.5 sccm of diluted $Ge_2H_6$ with a pressure range of $3x10^{-3}$mbar. A sudden decrease of temperature results in the rapid change in growth mode from a rapid step flow process at panel (a) to a hillock mode as in panel (b).

Similarly, a sudden change in precursor flow during growth while keeping the temperature constant results in an evolution from step-flow to 2D-nucleation (compare Fig.5a and 5b). At 0s in Fig.5b, the precursor flow is doubled from 1 to 2 sccm. The growth evolves continuously from step flow to a random 2-dimensional nucleation growth mode: small hillocks visible as bumps appear on the surface and subsequently disappear rapidly, characteristic of the birth and spread of numerous nuclei in the 2D nucleation model. The growth mode is further substantiated by dark and bright contrasts indicating a difference in thickness along the <11-20> zone axis. With increasing deposition rate, the 2D-nucleation rate increases and the size of 2D nuclei decreases. So, this regime results in a large increase of the overall {1-100} sidewall roughness.

It is of major importance to mention that the formation of hillocks on {1-100} Ge-2H facets is different from the strain-induced Stranski-Krastanov mode for which the growth is initially layer-by-layer until a critical thickness is reached at which the formation of 3D islands becomes energetically favoured. The GaAs-WZ and Ge-2H lattices are lattice matched. Thus, no strain-induced morphology change can be expected. In our experiments, there is no thickness threshold nor any presence of misfit dislocation. Moreover, the formation of hillocks can be initiated at any stage in the growth, also at the very beginning of the growth if the precursor flow is high at low temperature (SI4). Thus, this growth mode has a kinetic origin, as will be discussed below.

For a comprehensive review of the step-flow process see refs [27-31]. The step-flow is generally stable if the surface diffusion of the growth species is sufficiently high compared to the deposition rate. When surface diffusion is fast enough, adatoms can rapidly move to the most energetically favourable site on a step edge which progresses forward as described by the classical thermodynamic terrace-ledge-kink model developed by Kossel and Stranski.[32-33] If the precursor flow is increased, the deposition rate also increases and the system moves away from equilibrium. The adatoms do not reach the energetically favourable incorporation sites because of a too short diffusion length. This diffusion length becomes smaller than the distance between steps and 2D nucleation takes place on terraces leading to the hillock growth mode. Alternatively, decreasing the temperature while maintaining the deposition rate constant will reduce the mobility of adatoms. A step meandering may establish and finally a reduced surface diffusion length may also lead to random 2D-nucleation and thus surface roughening. In our experiments the growth mode of Ge-2H on m-plane GaAs-WZ can range from step-flow at high temperature, via unstable step-flow (step meandering) to 2D-nucleation stated by random hillock formation at low temperature and high deposition rate.

### 3.3 Formation mechanism of $I_3$-BSF during growth

Both $I_3$-BSF nucleation and the growth mode are tuned by the growth temperature and precursor flow. At first sight, one could suppose that $I_3$-BSFs formation is correlated with the induced roughness consecutive to destabilized step-flow. However, an initial surface roughness created by inhomogeneous thermal desorption of the GaAs NW does not cause the formation of defects, provided that the step flow mode is maintained and indeed flattens the surface (see SI5). Likewise, the high density of terraces on vicinal sidewalls (i.e. tapered sidewalls, making an angle with the m-plane) always formed on the surface of self-catalysed NWs does not play a noticeable role on the formation of $I_3$ defects (see SI3). In case of vicinal surfaces, step-flow flattens again the surface and $I_3$-BSFs are only formed in case of destabilised step-flow at high flow and low temperature. So, the surface roughness (craters or steps) is not a direct cause of $I_3$ formation and the relationship between roughness and $I_3$-BSF formation is thus not straightforward and must be considered as an indirect relationship.

Although the enhanced roughening during destabilised growth is not directly the cause of $I_3$ defect formation, it is obvious that $I_3$-BSFs and hillocks appear in the same growth condition regime and both phenomena are due to kinetics effects. In our *in situ* studies, no $I_3$ defect has been observed in case of step-flow. The more the step-flow is destabilised the higher the density of $I_3$ defects. Vice versa, the presence of numerous $I_3$-BSFs may successively worsen the surface roughness. We argue that the mechanisms involved in the chemisorption of adatoms on the surface not only guide the evolution of the surface morphology but also impact the formation of $I_3$ defects. In this paper, we assume and demonstrate that a lacking dangling bond may be at the origin of the $I_3$ defect and this is enhanced at high flow and low temperature. The cause of a missing dangling bond (vacancy, impurity, distortion…) is not yet established but discussed later.

In figure 7 (A-F), we propose a possible scenario to explain the formation of $I_3$-BSFs trying to account for all the observations. The specific triangular shape is of major importance indicating that the $I_3$ defect locally nucleates due to either a lacking dangling bond or the presence of an H atom, an unknown foreign atom or a vacancy in the Ge lattice that likewise prevents the chemisorption of the adatom at the right place. In fig. 7A the initial growth surface is seen in perspective view with an upward m-plane and a step-flow propagation in the <0001> direction from left to right. Based on the step-flow growth mode, the adsorbing surface is only the {1-100} surface and the Ge adatoms are forced to follow the stacking ruled by the {1-100} surface. We assume that the growth proceeds via step-rows along the <11-20> direction, and row-by-row in the <0001> direction. As discussed above, this growth mode enables to replicate a perfect 2H structure with the stacking of successive layers. The grown layer is in fact assumed to be constituted of 2 mono-atomic planes: each lower Ge atom is stacked and bonded with two upper Ge atoms.

In fig. 7B, at some point on the surface we place a "wrong atom" that hinders other atom to bond to it. This can also be a 'regular' Ge atom that is passivated by a hydrogen atom, as will be discussed below. Consequently, on the (1-100) surface an upward dangling bond is missing, preventing the chemisorption of the next upper Ge adatoms on the (1-100) surface. For the sake of simplification, we consider here the successive sticking of 3 Ge atoms that complete the tetrahedra on the surface. Consequently, a single atom vacancy in a lower plane of a layer will induce a double atom vacancy in the upper plane of the bi-atomic layer. The first Ge layer above this missing dangling bond will then display 3 vacancies as shown in fig. 7C. This configuration opens an empty space that forces two Ge atoms to stack in the <0001> direction rather than on the (1-100) surface. The opened $\{0002\}_{2H}$ basal plane corresponds to the $\{111\}_{3C}$ plane and thus will favour a cubic ABC stacking in the <0001> direction. So, a third Ge atom will sit in a faulted C position as shown in Fig. 7D. In fig. 7E, the next row of adatoms will continue to sit on the (1-100) surface except beside the faulted atom. To satisfy the <0001>-oriented dangling bond with the faulted Ge atom the neighbouring atom is forced to take a faulted C position. Elsewhere, the growth continues by step-flow on the (0001) surface. Therefore, on the first grown bi-atomic layer, the missing dangling bond has provoked the formation of a dumbbell of faulted atoms (Fig. 7F). This is the seed of the $I_3$-BSF. For the next layer, Ge ad-atoms again satisfy the bonds on the {1-100} surface via a step-flow mode. Successively grown layers reproduce the stacking fault (Fig. 7G and 7H). The number of faulted atoms increases by one on each grown layer and the faulted (0002) plane is bounded by a vacancy on each side (Fig. 7I). Following the proposed scenario, a triangular $I_3$-BSF is formed in an overall perfect 2H-Ge structure. The projections of the defect in various directions are shown in Fig. 8. They perfectly represent the observed TEM images of the defect (Fig.1). A slice of the faulted planes forming the $I_3$ defect viewed in the <0001> zone axis shows the arrangement of the triangular defect and the bounding vacancies. It is also important to mention the configuration of the $I_3$ defects when observed along the <11-20> lines.

This model is thought to be applicable for the 2D-growth of Ge-2H, Si-2H, as well as for the metastable WZ structures, as long as growth occurs on a non-polar m-plane facet. The formation of I₃-BSF is very likely in the polytype structures where the ZB structure is of lower formation energy than for the WZ one such as GaAs-WZ, GaP, GaN, InN, AlN. [6]

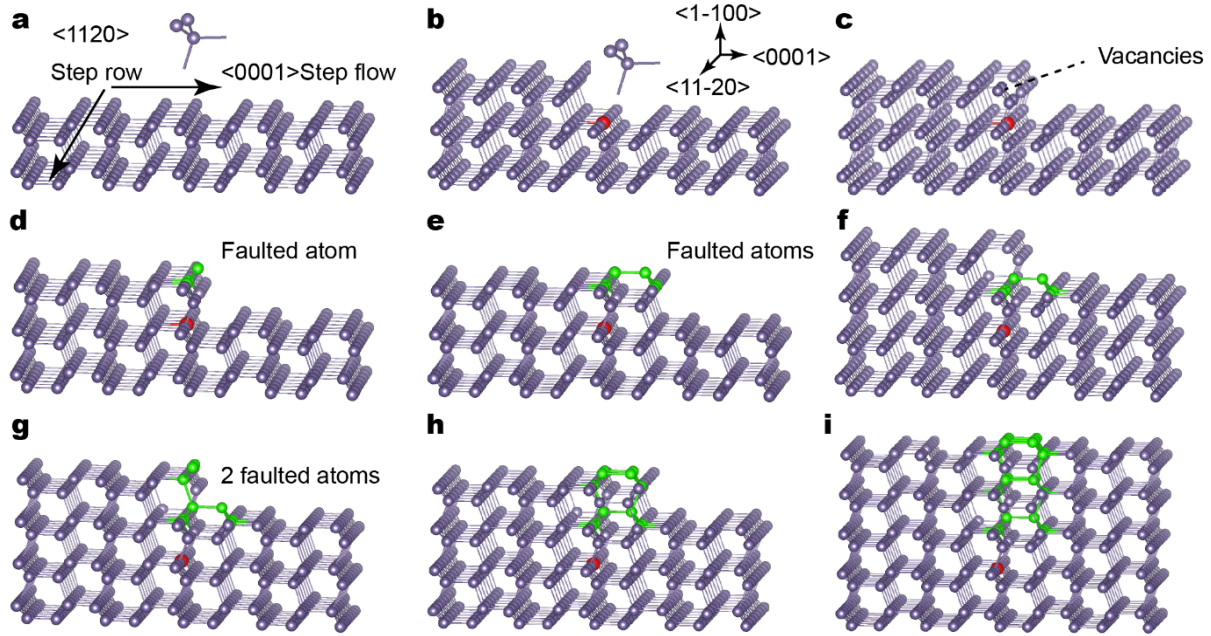

**Figure 7. Scenarios of *I*₃-BSF formation in Ge-2H :** The model shown in perspective view was drawn using vesta code. **a)** Initial perfect Ge-2H structure. The growth occurs upward on the (1-100) surface by addition of biatomic layers. Ge tetrahedra construct the surface by propagation along <11-20> oriented rows . The step flow progresses in the <0001> direction. **b)** An atom/impurity (in red) hinders a dangling bond. **c)** The lacking dangling bond prevents Ge chemisorption on the (1-100) surface leading to three vacancy positions. It opens locally a free surface with <0001> normal direction. **d)** At the vacancy location, Ge adatoms stick along the <0001> direction rather than on the (1-100) surface. Along <0001> direction, the stacking is the standard ABC cubic one. One Ge atom is in the faulted C place. **e)** The step flow advances along <0001>. On the following step, the bond to the faulted Ge atom forces a second Ge atom to a C faulted position. **f)** The growth continues by step-flow on the (0001) surface. A first biatomic layer has been completed. The missing dangling bond has provoked a dumbell of faulted atoms. **g)** A second layer is propagating on top of the first layer. Ge tetrahedra stack up on the (1-100) surface. On top of the faulted atom the tetrahedra is forced to rotate. Two more atoms are forced in the faulted C position. **h)** On top of the two lower faulted atoms tetrahedra replicate the fault. **i)** The further grown layer reproduces the stacking fault increasing the number of faulted atoms. For each additional layer on the (1-100) surface the *I*₃ BSF extends along <11-20>. There is one additional Ge faulted atom at each layer.

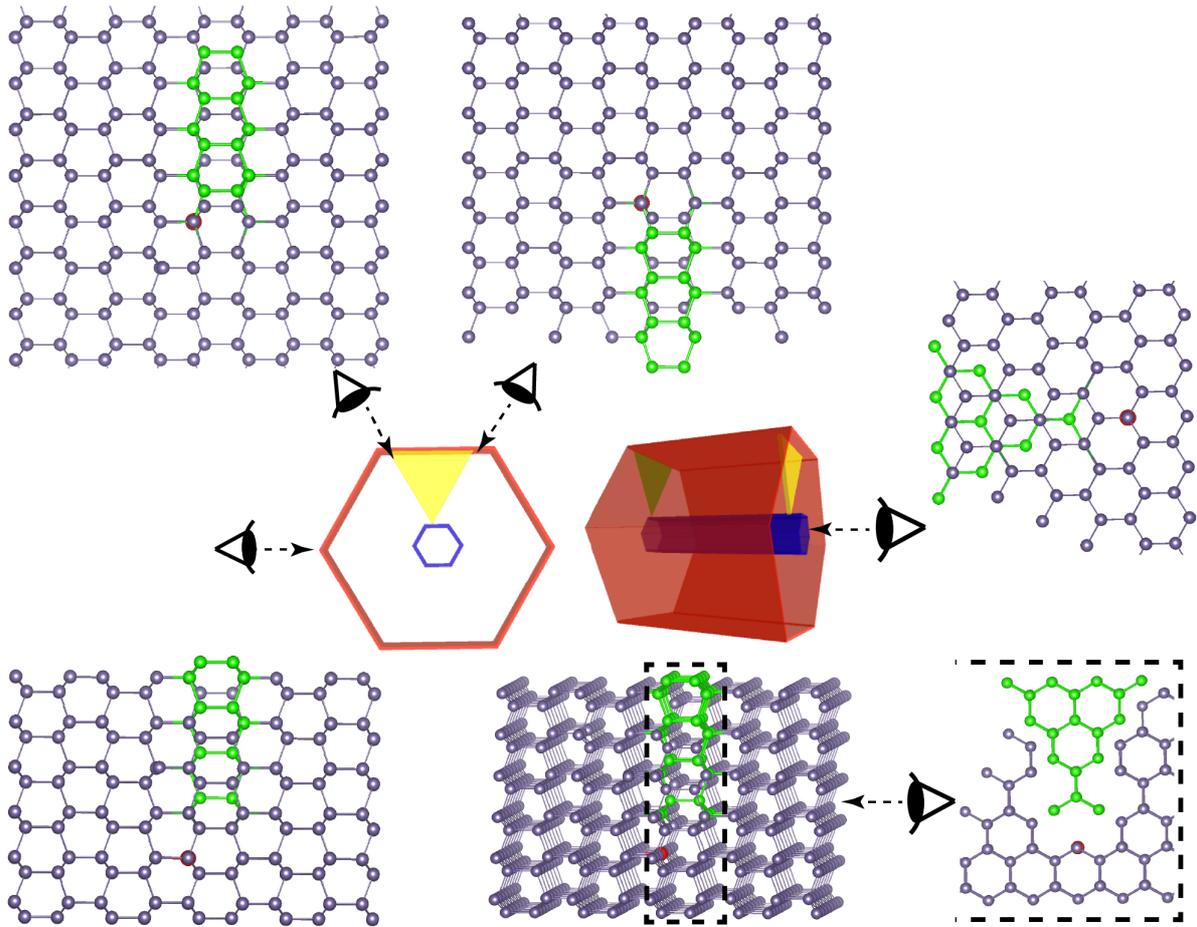

**Figure 8.** 2D-projection of atom positions of the $I_3$-BSF for various observation directions. Modeling is performed using VESTA software. The perfect 2*H* structure atoms are presented in purple and the $I_3$-BSF atoms are in green.

**Discussion on the origin of I$_3$-BSFs**

The proposed model is consistent with the *ex situ* and *in situ* observations. The following conclusions are of particular importance:

i) We observe a random nucleation at a discrete point and a triangular shape of the I$_3$ BSF. This may be explained by a single lacking dangling bond acting as a source for I$_3$ BSF formation. The triangle with <11-20> sides is the consequence of the layer-by-layer (1-100) stacking on top of the point source. There is an incremental increase of the number of faulted atoms with each layer.

ii) We do not see any influence of Au, Ga, As and Si contamination. It is important to note that Ga, As and Si have the same sp3 configuration as Ge and thus do not change the dangling bond configuration on the surface. The fact that those impurities seem not to have any influence on the defect nucleation is consistent with the increased I$_3$-density with higher

Ge$_2$H$_6$ flow while the contamination influence should be lower for higher Ge$_2$H$_6$ precursor flow. On the other hand, Au may act as liquid catalyst and it is highly mobile on the surface. Again, its possible influence would not be expected to increase with Ge$_2$H$_6$ flow.

iii) Indeed, the nucleation probability of I$_3$-BSFs relies on growth kinetics. It is unambiguously related to the growth conditions and probably to the growth kinetics as we will discuss below i.*e.* influenced by temperature and deposition rate. I$_3$-BSFs are associated with a destabilised step-flow. The more the step-flow is destabilized, the more the density of I$_3$-BSF increases. No defects are observed with a continuous step-flow growth mode when surface diffusion is high enough.

The last point requires further examination of the cause of step-flow destabilisation and the identification of the missing dangling bond at the origin of the nucleation of I$_3$-BSFs. This new configuration of the m-plane growth surface for Ge and Si requires a global study of dangling bonds and surface reconstruction. We believe that surface chemistry and particularly hydrogen coverage - being highly dependent on the growth temperature - may be the major cause. It is important to note that H atoms can passivate the dangling bonds while Ga, As and Si do not.

The presence of H on the growth surface is inherent to the CVD growth process and cannot be suppressed when using hydride precursors. H coverage during growth process affects diffusion processes and thus dramatically changes the growth kinetics and the surface morphology. [34-36] It is well documented that H coverage is influenced by both the temperature and the deposition flow [37, 38] and it plays a critical role on the quality of cubic Si and Ge films during CVD growth. [34, 39-44] The kinetics are governed by hydrides physisorption and H desorption which are thermally activated processes. The hydride sites are inactive for chemisorption unless H is released and leaves a dangling bond useful for an adatom or a hydride molecule. Thus, the sticking probability of new growth species depends on available active sites (dangling bonds) and surface mobility. Studies on CVD of cubic Si and Ge have shown that, at low temperature and high flux, the H desorption is the rate-limiting process while at high temperature and low flux the rate is limited by adsorption [37]. The hydrogen during CVD of Ge and Si epitaxy has two cumulative effects: firstly, H saturation reduces the adatom surface mobility by increasing the diffusion barrier and in turn provokes enhanced surface roughening, secondly, it saturates the step-edges, favouring island growth over step-flow growth.[45 - 46]

Thus, we assume that H passivation of dangling bond is the main cause of the observed morphological instabilities (step meandering) as well as roughening at low temperature or high flow rate. Thus far, no information is available regarding the H coverage of the Ge {1-100} surface. For cubic Si and Ge, the chemisorption mechanism is dependent on the crystallographic orientation.[47] For the {1-100} surface of hexagonal Si and Ge, it was reported that there are two dangling bonds per unit cell and the number of dangling bonds is not reduced upon reconstruction.[48] We can speculate that also for the m-plane, H saturation of dangling bonds may in some way reduce adatom mobility. The observed step-flow is consistent with the dangling bond diffusion proposed by Kuwahara *et al.* rather than a diffusion of adatoms.[49] Further study will be necessary to precisely determine the role of H in changing the growth modes of Ge-2H on the m-plane. Various experiments are planned to study the influence of H by comparing molecular beam epitaxy using atomic beam and CVD using hydrides gases but also introducing atomic hydrogen on the surface during growth.

## 4. Conclusion

*In situ* TEM observations provide direct evidence that the formation of unusual $I_3$-BSFs is associated with the instabilities of the step flow growth mode depending on deposition rate and thermal conditions. Therefore, the formation of $I_3$-BSFs may be avoided at high temperature providing the deposition flow is low enough to keep a continuous step-flow regime. In the specific case of Ge-2H/GaAs-WZ, the temperature should be as high as possible while considering the limit of 650°C imposed by the sublimation temperature of GaAs and the desorption rate of Ge atoms. We have emphasised the importance to understand the growth mechanisms to achieve defect-free Ge-2H growth on m-planes. Finally, the results provide us credible arguments to propose a model for the formation of $I_3$-BSF in the metastable 2H structures based on a missing dangling bond on the (1-100) surface that forces a Ge atom to sit on a faulted cubic position. The model accounts for the nucleation on discrete points and the resulting triangular shape of the $I_3$ defect. Any cause of reduction of surface diffusion leading to destabilisation of step-flow must be suppressed. For instance, hydrogen passivation must be further investigated.

## Experimental methods

In this study, we utilize Au-catalyzed GaAs-WZ nanowires as a template for the growth of Ge-2H shells in both *ex situ* and *in situ* TEM characterization experiments.

### *Ex situ growth*

For the *ex situ*, post-growth characterization experiments, the GaAs/Ge core/shell nanowires were grown in a low pressure (50 mbar) Aixtron close coupled shower head metal organic chemical vapor deposition (CCS-MOCVD) reactor. The growth of both the core and shell nanowires was performed at a total reactor flow of 8.2 standard liters per minute (slm) utilizing hydrogen ($H_2$) as the carrier gas. The GaAs nanowires were grown via catalyst-assisted growth following the Vapor-Liquid-Solid (VLS) mechanism utilizing gold (Au) catalyst seeds. The Au catalyst seeds were deposited in nano disks arrays arrangement on a GaAs (111)B substrate for the GaAs/Ge nanowires. For the GaAs nanowires, the growth template was annealed at a surface temperature of 569 °C (equivalent to thermocouple set temperature 635 °C) under an $AsH_3$ flow set to a molar fraction of $\chi_{AsH3} = 6.1 \times 10^{-3}$. Then, the WZ-GaAs nanowires growth was performed at a temperature of 615°C with TMGa and $AsH_3$ as material precursors set to molar fractions of $\chi_{TMGa}=1.9 \times 10^{-5}$, $\chi_{AsH3}=4.55\times10^{-5}$, respectively, achieving a total flux V/III ratio of 2.4. After the growth of the GaAs core nanowires, they are chemically treated with a diluted (1:10) potassium cyanide (KCN) aqueous solution, followed by a diluted ammonia solution ($NH_4OH$) step to remove residual GaAs oxide prior to subsequent Ge shell by MOVPE. Eventually, the GaAs nanowires were reintroduced in the MOVPE reactor and were used as a hexagonal material template and are overgrown with Ge. We used germane ($GeH_4$) (1% diluted) as a gas precursor for the Ge shell growth. The Ge shells were grown at substrate surface real-time monitored temperatures in the range of 405-565 °C (equivalent to 550-650 °C set temperature by a thermocouple) and at $GeH_4$ molar fraction of $\chi_{Ge}=8.5\times10^{-7}$ for a certain growth period according to the desired shell volume. It is worth mentioning that the growth of each material family (GaAs and Ge) has been performed utilizing a separate inner quartz kit and a substrate holder, *i.e.* susceptor in the MOCVD growth chamber to avoid materials cross-contamination. The details of the GaAs/Ge core/shell nanowires growth is summarized in supporting information SI1.

In this study, we have investigated multiple surface treatment conditions for the GaAs core nanowires templates prior to the Hex-Ge shell growth as explained in SI1. The following three different GaAs nanowire templates have been utilized to mimic three different interface qualities to study their possible impact on the formation of the $I_3$ BSFs in the Hex-Ge shell:

(1) As-grown GaAs core nanowires: after growth, the as-grown, Au-catalyzed GaAs nanowires are stored in a glove box under an ambient of nitrogen ($N_2$) to prevent surface oxidation and contamination. The utilization of the as-grown GaAs nanowires is meant to avoid any possible contamination or different surface termination that may occur during the chemical treatment of the nanowires' surface to remove the Au particles. Hence, this type of GaAs cores template is referred to, hereafter, as 'untreated as-grown' core surface, yet with Au particles. It is worth mentioning that after the chemical etching step of Au particles, APT experiments have been performed on the etched GaAs core nanowires and the as-grown GaAs/Ge core/shell nanowires structures with Au present at the top, and no Au traces have been detected as shown in Refs. [9, 11]. Hence, the concentration of Au impurities -if any present-is beyond the APT detection limit. So, we can neglect the influence of Au on the crystal structure of Hex-Ge.

*(2)* Chemically treated GaAs core nanowire: in this template, the as-grown GaAs nanowires are chemically treated with a diluted cyanide (KCN in deionized water) solution to etch away the Au particles. They are then treated with diluted ammonia ($NH_4OH$ in deionized water) to remove residual GaAs oxides. This template will be referred to as a 'chemically treated' core surface.

(3) Annealed GaAs core nanowires: to obtain a clean core surface, free of potential wet chemical contaminants, after the Au etching step, the GaAs cores have been annealed at a high surface temperature 613°C under high $AsH_3$ pressure to desorb any existing contaminants after the annealing step. This template will be referred to as the 'annealed' core template.

### *In situ growth*

The growth of GaAs NWs and the epitaxy of Ge shells are observed *in situ* using a Cs-corrected Titan environmental TEM (ETEM) operated at 300 kV. Movies are recorded using a Gatan US1000 camera at a rate of 4 fps. The images in the article are snapshots extracted from the videos.

We use a Protochips FUSION sample holder with heating chips featuring a SiC membrane with 9 holes of 10 μm diameter and providing uniform heating in the observation area. Because of the polycrystalline nature of the SiC membrane, NWs are grown in arbitrary directions on the substrate. Some of them are anchored on the edge of a hole and may be freely suspended to be observable in a hole. Although the FUSION chips are calibrated by the manufacturer, we observe huge fluctuations of temperature from one chip to another. So, we rather empirically estimate a range of temperature using the melting point of Au on Si at 320°C and looking at the tapering of Au catalysed Ge nanowires (*i.e* no tapering under 400°C and increasing tapering with temperature).

The VLS growth of GaAs NWs is performed using trimethylgallium (TMGa) and tertiarybutylarsine (TBAs) and the Ge shell is grown using digermane ($Ge_2H_6$) diluted in $H_2$ (10% by volume). TMGa, TBAs and $Ge_2H_6$ gases are installed in a gas cabinet. Diluted $Ge_2H_6$ is stored in a cylinder at a pressure of 5 bars and the flow is controlled by a MFC (mass flow controller). The TMGa and the TBAs liquids are stored in bubblers with a vapor pressure at 20°C of 246 and 166 mbar respectively; these low vapor pressures are sufficient to regulate their flow rates by conventional MFCs without the need for an additional carrier gas. The mixture is then delivered to the microscope column by an injector connected near the pole piece. During growth experiments, the pressure near the pole piece was measured by a Pirani pressure gauge. An additional compensated vibrational home-made pumping system is mounted on the column; this system is based on two identical bellows from either side of the additional turbo pump (80l/s). We can thus tune the column pressure (by setting the rotation rate or closing the valve) for a given precursor flow.

For the growth of GaAs nanowires, Au nanoparticles are deposited by evaporation on the heating chip. The *in situ* growth of the GaAs template is described in SI2 both for Au and self-catalysed GaAs NWs. The Au catalyst is useful to control the ZB to WZ crystal phase transition, the diameter of the W part (around 10 to 30 nm) and the morphology with very flat sidewall surfaces. In the case of self-catalyzed GaAs NWs, the lateral overgrowth is enhanced and we obtain very large NWs with inevitably high cut-off vicinal sidewall surfaces. Moreover, it is quite difficult to maintain the Ga droplet smaller than 50 nm in the WZ mode growth which is detrimental for high resolution analyses. These NWs may however be used to understand the influence of the initial surface roughness or steps as well as potential effects of Au contamination on the Ge shell (in SI3).

After the growth of GaAs NWs, organometallics are simultaneously stopped and diluted $Ge_2H_6$ is injected.

**Supporting videos**

Supporting videos are available to download from https://doi.org/10.5281/zenodo.5665207

**SV1** : Growth of GaAs nanowire at 500°C transition from ZB to W structure by changing the V/III ratio from 2 to 15 at a pressure around $2x10^{-3}$ mbar.

**SV2** : Perfect epitaxial growth of Ge-2H at 450°C with 1sccm $Ge_2H_6$ and a pressure around $2.10^{-3}$ mbar. Growth rate is 1.5 Å/min.

**SV3** : Same NW of SV2. Growth of Ge-2H at 450°C with 2.5 sccm $Ge_2H_6$ at a pressure range around $5.3x10^{-3}$ mbar. The growth rate is of 20 Å/min. Increasing the flow of $Ge_2H_6$ induces hillock formation. The destabilisation is associated with I3-BSF formation and the density increases with growth time.

**SV4** : Step-flow viewed with <11-20> zone axis. The growth is performed at around 500°C, 1sccm of diluted $Ge_2H_6$ and P= $2.5x10^{-3}$ mbar.

**SV5** : Tilted view of the step-flow growth at 550°C and 1.2 sccm $Ge_2H_6$ with P= $1 \times 10^{-3}$ mbar.

**SV6** : Enhancement of step meandering. Starting at 450°C with 1 sccm of $Ge_2H_6$ , the temperature is reduced to 420°C at 95 s.

**SV7** : Same growth as in SV4 performed at 500°C but with increased deposition flow to 2 sccm of Ge2H6 at P= $4.3 x10^{-3}$ mbar. The increased flow destabilises the step-flow and leads to the formation of hillocks and few BSFs.

**SV8** : VESTA 3D-vizualization of a $I_3$ defect

**Supporting information**
Supplementary info can be found in a separate file.

**Author contribution**
TU/e part : E.F. has performed the growth of the nanowires, M.A.V. has done the TEM experiments. E.F., M.V, W.H.J.P., M.A.V. and E.P.A.M.B. have analysed the data. M.A.V. and E.P.A.M.B. have co-supervised the ex-situ part of the project.
C2N part : L.V, C.R, D.B. designed and conducted the in situ experiments with assistance of F.P. and I. F.L.V. analysed the data from in situ experiments, developed the model and wrote the manuscript. All authors discussed the results and commented on the manuscript.
All authors have given approval to the final version of the manuscript. E.P.A.M. B and L.V acquired the research funding that supported this study.

**Acknowledgements**
This project has received funding from the European Union's Horizon 2020 research and innovation program under grant agreement No 735008 (SiLAS) and No 964191 (OptoSilicon) and the French National Research Agency (ANR) under the grant ANR-17-CE030-0014-01 (HEXSIGE). We acknowledge Solliance, a solar energy R&D initiative of ECN, TNO, Holst, TU/e, imec, orschungszentrum Jülich, and the Dutch province of Noord-Brabant for funding the TEM facility.
We acknowledge the ANR for funding the NANOMAX ETEM through the TEMPOS grant (10-EQPX-0050)). We acknowledge Odile Stephan leader of TEMPOS and Jean Luc Maurice manager of NANOMAX. We wish to particularly acknowledge Ileana Florea for the technical assistance on the NANOMAX facility and her great availability during experiments. Thanks are due to the CIMEX at École polytechnique (Palaiseau, France) for hosting NANOMAX microscope.

**Conflict of interest**
The authors declare no competing financial interest

**Data availability**

The raw and processed data required to reproduce these findings are available to download from https://doi.org/10.5281/zenodo.5602171